\def\draftversion{false}

\RequirePackage{ifthen}
\ifthenelse{\equal{\draftversion}{true}}{
  \documentclass[prb,galley,showpacs,preprintnumbers,citeautoscript,
      amsmath,amssymb]{revtex4}
}{
  \documentclass[citeautoscript,floatfix,aps,prb,twocolumn,
      superscriptaddress]{revtex4-2}
}

\usepackage{amsmath}
\usepackage{graphicx}
\usepackage{epstopdf}
\usepackage{natbib}
\usepackage{array}
\usepackage{subcaption}
\usepackage[utf8]{inputenc}
\usepackage{xspace}
\usepackage{cleveref}
\usepackage{multirow}
\usepackage{changes}

\usepackage{dcolumn}
\usepackage{bm}

\usepackage{soul}  

\newcommand\thefontsize{The current font size is: \f@size pt}
\newcommand\crpau{$U_{\text{cRPA}}$}

\ifthenelse{\equal{\draftversion}{true}}{
  \marginparwidth 2.7in
  \marginparsep 0.5in
  \newcounter{comm} 
  \def\commnext{\stepcounter{comm}}
  \def\commtext{{\bf\color{blue}[\arabic{comm}]}}
  \def\commmar{{\bf\color{blue}[\arabic{comm}]}}
  \def\dvm#1{\commnext\marginpar{\small DV\commmar: #1}\commtext}
  \def\cdm#1{\commnext\marginpar{\small CED\commmar: #1}\commtext}
  \def\msm#1{\commnext\marginpar{\small MS\commmar: #1}\commtext}
  \def\mlab#1{\marginpar{\small\bf #1}}
  
}{
  \def\dvm#1{}
  \def\cdm#1{}
  \def\msm#1{}
  \def\mlab#1{}
  
}

\newcommand{\cubic}{$Pm\overline{3}m$\xspace}
\newcommand{\smo}{SrMoO$_3$}
\newcommand{\pmo}{PbMoO$_3$}
\newcommand{\lmo}{LaMoO$_3$}

\begin{document}
\title{First-principles study of the electronic, magnetic, and crystal structure of perovskite molybdates}
\author{Jeremy Lee-Hand}
\affiliation{Department of Physics and Astronomy,
             Stony Brook University,
             Stony Brook, New York, 11794-3800, USA}
\author{Alexander Hampel}
\affiliation{Center for Computational Quantum Physics,
             Flatiron Institute,
             162 5th Avenue, New York, New York 10010, USA.}
\author{Cyrus E. Dreyer}
\affiliation{Department of Physics and Astronomy,
             Stony Brook University,
             Stony Brook, New York, 11794-3800, USA}
\affiliation{Center for Computational Quantum Physics,
             Flatiron Institute,
             162 5th Avenue, New York, New York 10010, USA.}
\date{\today}

\begin{abstract}
The molybdate oxides SrMoO$_3$, PbMoO$_3$, and LaMoO$_3$ are a class of metallic perovskites that exhibit interesting properties including high mobility, and unusual resistivity behavior. We use first-principles methods based on density functional theory (DFT) to explore the electronic, crystal, and magnetic structure of these materials. To account for the electron correlations in the partially-filled Mo $4d$ shell, a local Hubbard $U$ interaction is included. The value of $U$ is estimated via the constrained random-phase approximation approach, and the dependence of the results on the choice of $U$ are explored. For all materials, generalized-gradient approximation DFT+$U$ predicts a metal with an orthorhombic, antiferromagnetic structure. For LaMoO$_3$, the $Pnma$ space group is the most stable, while for SrMoO$_3$ and PbMoO$_3$, the $Imma$ and $Pnma$ structures are close in energy. The $R_4^+$ octahedral rotations for SrMoO$_3$ and PbMoO$_3$ are found to be overestimated compared to the experimental low-temperature structure.
\end{abstract}

\maketitle

\section{Introduction}
Perovskite oxides, which have the chemical formula $AB$O$_3$, have received much attention due to the range of properties they exhibit, including ferroelectricity~\cite{Dawber2005,Nova2019,Khomskii2009}, superconductivity~\cite{Collignon2019,RevModPhys.66.763}, various magnetic orders~\cite{Callaghan1966, Zhang2015}, and catalytic activity~\cite{Ishihara2017}. Three important reasons for the diverse properties in this material system are: the diversity of elements that make up perovskites \cite{Mitchell2002}; the  fact that perovskite oxides have been found to crystallize in a range of different structures based on distortions of the cubic (\cubic) motif, including those with rhombohedral, orthorhombic, and hexagonal symmetry~\cite{Lufaso2001}; and the fact that high quality heterostructures and thin films can be grown, combining the functional properties of individual perovskite oxides and resulting in novel phenomena \cite{Tsymbal20121}.

Molybdate oxides such as SrMoO$_3$ (SMO), PbMoO$_3$ (PMO) and LaMoO$_3$ (LMO), are a class of the aforementioned perovskites with interesting properties, but relatively little experimental and theoretical characterization. SMO has been found to have the highest known electrical conductivity of all perovskite oxides~\cite{Nagai2005}, which is promising for electrode applications in oxide devices. Low levels of alloying with Cr also drives a ferromagnetic (FM) transition~\cite{Zhao2007}. Furthermore, PMO has been reported to have unusual electrical properties, exhibiting a sub-linear resistivity with a low-temperature peak that does not appear to be related to a structural transition~\cite{Takatsu2017}. Very little is known about LMO, but similar perovskites (such as LaCoO$_3$) display very high electrical conductivity \cite{Senaris-Rodriguez1995}, and perovskite oxides with La in the $A$-site have been identified as promising solid oxide fuel cells cathodes~\cite{Huang1998}. 

In this work, we present a systematic exploration of the atomic, magnetic and electronic structure of SMO, PMO, and LMO, based on density functional theory plus Hubbard $U$ (DFT+$U$) calculations. We select SMO and PMO for study because of their interesting experimental properties (described further in Sec.~\ref{sec:exp}). LMO is included because, as we show in Sec.~\ref{sec:results}, the different formal charge of the $A$-site cation with respect to SMO and PMO results in distinct properties. This contrast assists in understanding (and provides an avenue for tuning) the behavior of SMO and PMO. We aim to address inconsistencies in previous first-principles calculations and stimulate further experimental work on these materials. 

\begin{figure}
    \centering
    \includegraphics[width=0.95\linewidth]{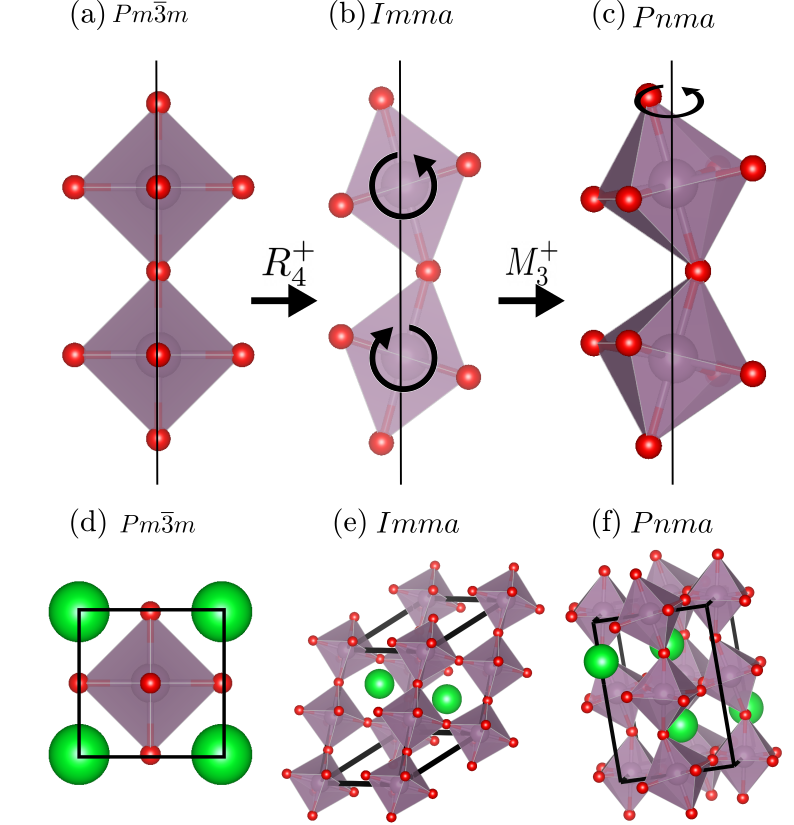}
    \caption{Different crystal structures related by structural distortions of MoO$_6$ octohedra. (a) The undistorted octahedra of the cubic \cubic structure; (b) out-of-phase rotations produce the $Imma$ structure, and (c) an additional in-phase rotation results in $Pnma$. (d), (e) and (f) show the unit cells of \cubic{}, $Imma$, and $Pnma$, respectively, used in this study. Green atom correspond to Sr/Pb/La, purple atoms are Mo, and red atoms are O.}
    \label{fig:crystal_structures}
\end{figure}

We explore the experimentally reported cubic $Pm\overline{3}m$ and orthorhombic $Imma$ structures, as well as the $Pnma$ structure (all shown in Fig.~\ref{fig:crystal_structures}), which have Glazer ~\cite{Glazer1972} notations ($a^0 a^0 a^0$), ($a^0 b^- b^-$), and ($a^- b^+ a^-$) respectively. Though not observed experimentally for the molybdates, the $Pnma$ structure is often reported as the lowest energy structure of similar perovskite oxides which also take on the $Pm\overline{3}m$ and $Imma$ structures at different temperatures. We find that, when an onsite $U$ on the Mo $4d$ states is included in our calculations, all of these materials are predicted to be metallic with an orthorhombic crystal structure and antiferromagnetic (AFM) ordering. However the details of the structure for SMO and PMO, including the energy difference between $Imma$ and $Pnma$ and the magnitude of the octahedral tilts, are sensitive to the treatment of the correlations (i.e., the magnitude of $U$) and the magnetic structure. We will explain this sensitivity via analysis of the electronic structure and phonons in these materials, and comparison with the case of LMO which does not show this sensitivity.

\begin{figure}
    \centering
    \includegraphics{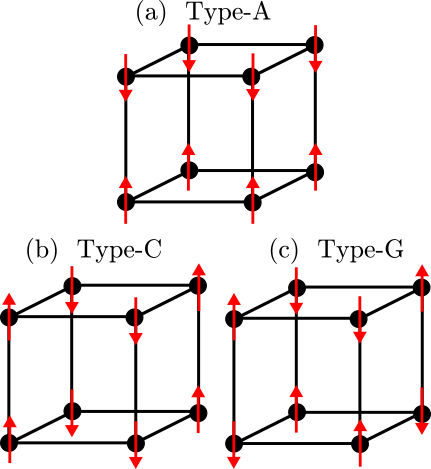}
    \caption{Antiferromagnetic orderings explored in this work: (a) type-A, (b) type-C, and (c) type-G. Black dots indicate Mo atoms, red arrows indicate spins.}
    \label{fig:magnetic_structures}
\end{figure}

The paper is organized as follows. In Sec.~\ref{sec:exp} we summarize the previous experimental and theoretical understanding of these materials. In Sec.~\ref{sec:comp_detail} we describe the computational methods and parameters that are used in this work. Sec.~\ref{sec:results} gives the electronic, atomic, and magnetic results of our calculations. A discussion of our results follows in Sec.~\ref{sec:discuss} with a comparison of similar perovskites and a simple analysis of structures predicted by the tolerance factor. Last, in Sec.~\ref{sec:conclusion} we conclude and suggest future directions for work in these materials.

\section{Experimental and previous theoretical work \label{sec:exp}}

Neutron diffraction studies by ~\citet{Macquart2010} observed that SMO has a $Pm\overline{3}m$ structure at room temperature (RT), and undergoes transitions to $I4/mcm$ at 266 K and $Imma$ at 125 K. Cubic structures have also been observed in SMO in RT X-ray diffraction experiments \cite{Cascos2016, Hopper2016}. Neutron and X-ray diffraction experiments by ~\citet{Takatsu2017} in PMO have shown similar results to SMO with a $Pm\overline{3}m$ structure at RT and a $Imma$ structure at 5 K. An attempt has been made to synthesize LMO~\cite{Figen2014}, but it did not yield a crystalline phase with 1-1-3 stoichiometry.

Resistivity measurements of SMO~\cite{Brixner1960, Nagai2005} and PMO~\cite{Takatsu2017} show metallic behavior down to low temperatures. In SMO, diffraction experiments show no evidence of magnetic order down to 5~K \cite{Macquart2010}, and another synthesis by ~\citet{Ikeda2000} shows susceptibility indicative of a Pauli paramagnet between 2 and 300 K with some Curie-like increases below 20 K. In PMO, ~\citet{Zhao2017} measure a Curie-Weiss-like susceptibility with a negative $T_c$ suggesting a paramagnetic state.

DFT calculations in the literature show no conclusive picture of the magnetic structure, and often make an assumption about the ground state crystal structure. Calculations of SMO assuming the $Pm\overline{3}m$ structure with generalized-gradient approximation (GGA) functionals found no spontaneous magnetization~\cite{Somia2019}. Local spin-density approximation (LSDA) calculations also found no magnetization in either the $Pm\overline{3}m$ and $Imma$ structures, with the application of a Hubbard $U$ for $0 \leq U \leq 6$~eV \cite{Zhu2012}. Other authors have reported an antiferromagnetic ground state in $Pm\overline{3}m$ with $U = 2$~eV \cite{Tariq2018} using GGA. In PMO, first-principles calculations of $Pm\overline{3}m$ report both non-magnetic~\cite{Dar2017} ($U = 2.5$~eV) and antiferromagnetic~\cite{Tariq2018} ($U = 0$~eV) magnetic structures. Finally, $Pm\overline{3}m$ and $Imma$ calculations of PMO report a nonmagnetic to ferromagnetic transition at $U = 3.8$~eV and $U = 3.5$~eV respectively and that the $Imma$ is lower in energy for a variation of $U$ from $0$ to $6$~eV~\cite{Nazir2017}; however, no antiferromagnetic structures of PMO were explored.

\section{Computational details}
\label{sec:comp_detail}

Bulk DFT calculations for SMO, PMO, and LMO were performed to obtain energies and electronic structure of the $Pm\overline{3}m$, $Imma$ and $Pnma$ structures with nonmagnetic (NM), ferromagnetic (FM), and antiferromagnetic type-A (AFM-A), type-C (AFM-C), and type-G (AFM-G) configurations (see Fig.~\ref{fig:magnetic_structures}).
Phonon calculations were performed for the NM case, in the \cubic, $Imma$ and $Pnma$ crystal structures.

\subsection{DFT parameters}
All calculations were performed using the VASP code~\cite{Kresse1993,Kresse:1996kl,Kresse:1999dk} with projector-augmented wave (PAW) potentials to describe the core electrons \cite{Blochl1994}. For each of the atoms, the explicit PAW projectors used to treat the valence states were: Sr ($4s,5s,4p,4d$), Pb ($5d,6s,6p,5f$), La ($5s,6s,5p,5d,4f$), Mo ($4s, 5s, 4p, 4d, 4f$), and O ($2s, 2p, 3d$). The GGA functional of Perdew, Burke, and Ernzerhof (PBE)~\cite{Perdew1996} was used for the exchange-correlation, and the DFT + Hubbard $U$~\cite{Lichtenstein1995} method was used to take into account the electron-electron interactions in the Mo $4d$ orbitals. Comparisons were also made using the hybrid functional of Heyd, Scuseria, and Ernzerhof (HSE)~\cite{Heyd2003, Heyd2006}.

Methfessel-Paxton smearing~\cite{Methfessel1989} of order two was used, with a smearing parameter of $\sigma = 0.05$~eV determined via convergence study of the total energy down to approximately $\pm 2$ meV. Convergence of the self-consistent field steps was set to $10^{-8}$~eV. A plane wave cutoff of $550$~eV was used in all calculations. Calculated total energies were found to be particularly sensitive to the plane wave cutoff, and careful convergence was required.
The $k$-point meshes used to sample the Brillouin zone were scaled to maintain consistent density for the different unit cell sizes of $Pm\overline{3}m$, $Imma$ and $Pnma$ [see Fig.~\ref{fig:crystal_structures}(d)-(f)], as well as the supercells necessary to represent the antiferromagnetic order. As a base case the $Pm\overline{3}m$ NM calculation ($5$ atoms per unit cell) were performed using a $16 \times 16 \times 16$ mesh.

Phonon calculations were performed using the finite-difference method of Refs.~\onlinecite{Lloydwilliams2015} and \onlinecite{Monserrat2018}, using forces calculated from VASP. A $2 \times 2 \times 2$ course $q$ grid was used for $Pm\overline{3}m$ and $3 \times 3 \times 3$ for $Pnma$ calculations and $Imma$ calculations for better resolution of soft modes. 

To determine octahedral-rotation-mode amplitudes, we performed a symmetry-adapted distortion mode analysis~\cite{PerezMato:2010ix} using the ISODISTORT software~\cite{Campbell2006}.

\subsection{Constrained RPA Calculation of $U$}
\label{sec:crpa}

\begin{table}
\centering
\caption{Values for the spherically averaged Slater parameter $F_0$ and exchange coupling $J=(F_2+F_4)/14$ obtained via cRPA calculations. The Wannierization window around the Fermi level and orbitals included (in addition to Mo $4d$) are also specified. All energy units are eV. }\label{tab:crpa}
\begin{ruledtabular}
\begin{tabular}{c | c c c c c }
& Window & Orbitals incl. & $U_{\text{cRPA}}\equiv F_0$ &  $J$ \\
 \hline
 SrMoO$_3$ & $-7.0$ to 2.1 & O $2p$, Sr $3d$ & 2.31 & 0.70 \\
 PbMoO$_3$ &$-7.3$ to 2.5& O $2p$, Pb $6d$ $6p$ & 1.84 & 0.65 \\
 LaMoO$_3$ & $-10.0$ to 5.0&O $2p$, La $4f$ & 1.78 & 0.65 
\end{tabular}
\end{ruledtabular}
\end{table}

To determine a value for the Hubbard $U$, we used the constrained random phase approximation (cRPA) \cite{Aryasetiawan2004} as implemented in the VASP code~\cite{Kaltak2015}, which allows the calculation of the effective partially screened Coulomb interaction by separating the electronic structure into a subspace near the Fermi level and the rest of the system. Formally, this means the separation of the total electronic polarizability $P = P_{\text{sub}} + P_{\text{rest}}$ where $P_{\text{sub}}$ is the polarizability for the correlated subspace (in our case, the Mo $4d$ orbitals) and $P_{\text{rest}}$ is for the rest of the system. From this, the screened  interaction tensor can be calculated in a local basis from the bare Coulomb interaction tensor $\bm{V}$, as $\bm{U}(\omega) = \bm{V}/[1 - \bm{V}P_{\text{rest}}(\omega)]$. Here, we limit ourselves to the static limit $\bm{U}(\omega = 0)$ of the screened interaction.

Since the DFT+$U$ implementation that we use in VASP is rotationally invariant \cite{Lichtenstein1995}, we perform a spherical average of the full four-index screened-Coulomb interaction tensor $U_{ijkl}$ to obtain the $F_0$ Slater parameter, and its corresponding exchange interaction parameter $J=(F_2+F_4)/14$ assuming $F_4/F_2=0.625$ (see Table \ref{tab:crpa}). 

To construct a well-localized correlated subspace for cRPA~\cite{Vaugier2012}, maximally localized Wannier functions were generated for all states in a wide energy window around the Fermi level using the Wannier90 package~\cite{Mostofi2014}. The NM state of the cubic materials were used for these calculations. As we will discuss in Sec.~\ref{sec:elec_struct}, these states involve hybridization between the Mo $4d$ orbitals (that we would like to target for our subspace), and O $2p$ or $A$-site orbitals. Including these other orbitals in the Wannierization, though excluding them for the calculation of $P_{\text{sub}}$, allows us to effectively disentangle the Mo $4d$ for our correlated subspace~\cite{Kaltak2015}.
For SMO, the O $2p$, Sr $3d$ and Mo $4d$ states were included in the Wannierization.
In addition to the O $2p$ and Mo $4d$ states, the Pb $6d$ and $6p$ states at the Fermi level were included for PMO, and the La $4f$ states just above the Fermi level for LMO. We found that the diagonal elements $U_{iiii}$ differ by less than 5\% between the Mo $4d$ basis orbitals, suggesting that the assumption of spherical interaction is well founded. 

We give the results for F$_0$, and $J$ in Table \ref{tab:crpa}. For all materials, the relatively low magnitudes of $F_0$ indicate strong screening of the Mo $4d$ orbitals by the states around the Fermi level, which is in good agreement for SMO with a previous recent cRPA study~\cite{Petocchi2020}. We see that the additional Pb-related states at the Fermi level for PMO and La $4f$ states for LMO result in further screening of the Coulomb interaction in the Mo $4d$ manifold. 

While our cRPA calculations give us crucial guidance as to the appropriate $U$ for the molybdate systems, these values are not guaranteed to be quantitatively accurate for all materials properties \cite{Kimber2009,Seth2017,Honerkamp2018}. Thus, in Sec.~\ref{sec:results} we will determine and discuss the sensitivity of our results to the chosen $U$ value, indicating where substantive changes may occur for slightly different choices of $U$. In addition, to gain insight into the driving forces behind the magnetic and structural properties, we show the general behavior of the electronic and crystal structure as a function of $U$ in a large range between 0 and 5 eV. The cRPA suggested values are indicated in all cases as $U_{\text{cRPA}}\equiv F_0$. For all materials, we use $J=0.7$.

\section{Results}
\label{sec:results}

\subsection{Electronic structure of cubic molybdates}
\label{sec:elec_struct}

\begin{figure*}
    \centering
    \begin{subfigure}[b]{0.5\textwidth}
        \centering
        \includegraphics[width=\linewidth]{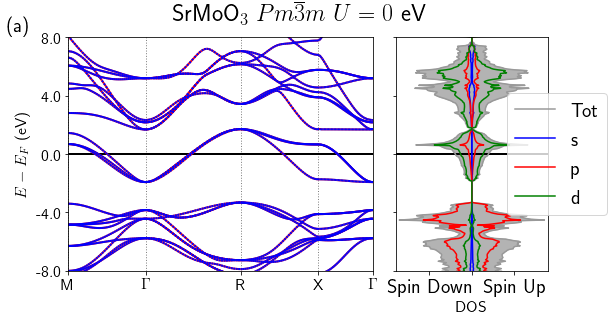}
        \label{fig:sr_band0}
    \end{subfigure}%
    \begin{subfigure}[b]{0.5\textwidth}
        \centering
        \includegraphics[width=\linewidth]{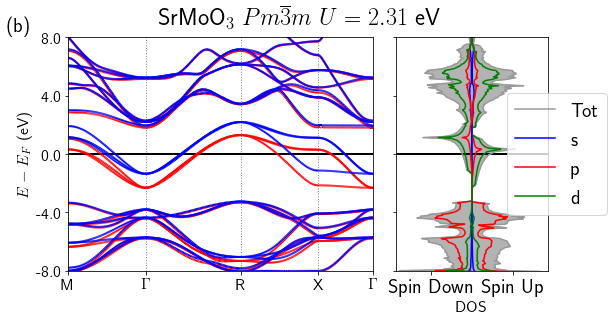}
        \label{fig:sr_band3}
    \end{subfigure}
    \\
    \begin{subfigure}[b]{0.5\textwidth}
        \centering
        \includegraphics[width=\linewidth]{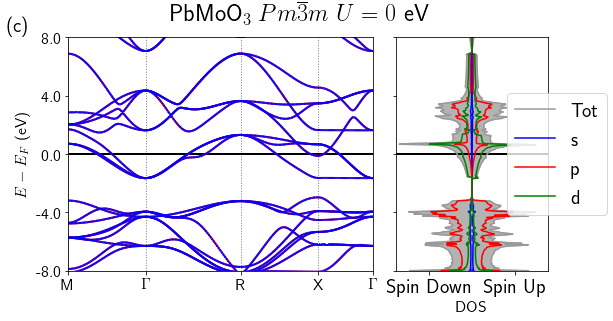}
        \label{fig:pb_band0}
    \end{subfigure}%
    \begin{subfigure}[b]{0.5\textwidth}
        \centering
        \includegraphics[width=\linewidth]{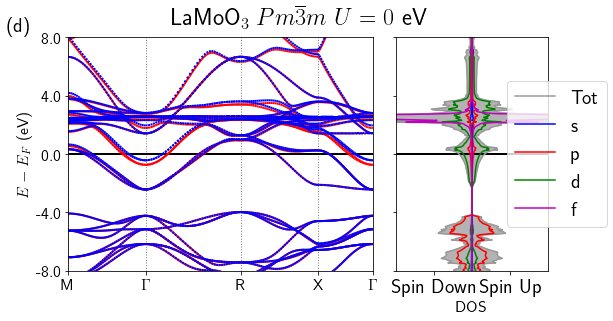}
        \label{fig:la_band0}
    \end{subfigure}
    \\    
    \caption{Band structure and density of states (DOS) for \cubic{} molybdates calculated with PBE+$U$, where $U$ is applied to the molybdenum $d$ electrons. (a) is SrMoO$_3$ with $U = 0$~eV, (b) is SrMoO$_3$ for $U = U_{cRPA}$~eV, (c) is PbMoO$_3$, $U=0$, and (d) is LaMoO$_3$ $U=0$. Note, that LaMoO$_3$ also has $U = 5$~eV applied to the $f$ electrons. Blue bands are spin up, red are spin down. }
    \label{fig:band}
\end{figure*}

To illustrate the basic elements of the electronic structure of these materials, we first focus on the cubic phase of the molybdates; in Figs.~\ref{fig:band}(a), (c), and (d), we plot the electronic band structures and density of states (DOS) for the cubic structures using $U=0$. In these cases, the bands are close to being spin degenerate, and there is no net magnetic moment. The electronic structure is similar between the materials, with the molybdenum $4d$ $t_{2g}$ states (green DOS) at the Fermi level, hybridized with oxygen $2p$ states (red DOS). The plateau along the $X - M - \Gamma$ line produces a van Hove singularity (vHS) in the DOS at about 1~eV above the Fermi level for SMO and PMO; for LMO,  the Fermi level is  only 0.3~eV below the vHS. This difference between SMO/PMO and LMO will affect the magnetic properties, as we will discuss in Sec.~\ref{sec:mag_struc}.

In Fig.~\ref{fig:band}(b) we show the band structure and DOS for SMO with $U=3$~eV. We see that including $U$ splits the spin up and spin down manifolds of the $t_{2g}$ orbitals, resulting in a net moment. We find that all three materials remain metallic (in agreement with experiment for SMO and PMO \cite{Brixner1960,Nova2019, Tiwari2015}) for the range of $U$ values studied here ($0\leq U \leq 5$~eV). 

For the case of LMO, PBE places the empty lanthanum $4f$ states quite close to the Fermi level. Since we do not expect that GGA will place the $f$ states correctly, we apply $U=5$~eV on the La $4f$ states to shift these states upward in energy [Fig.~\ref{fig:band}(d)]. We show in Appendix~\ref{sec:la_f_states} that adding this $U$  reproduces the DOS that is calculated with the HSE functional (Fig. \ref{fig:la_f_dos}). Though the application of this additional Hubbard $U$ does not change our results qualitatively (see Appendix \ref{sec:la_f_states}), we expect that it provides a more accurate picture of the electronic structure, and we will include it for LMO in all further results.

\subsection{Dynamic instability of \cubic{} \label{sec:phonons}}

\begin{figure*}
    \centering
    \includegraphics[width=\linewidth]{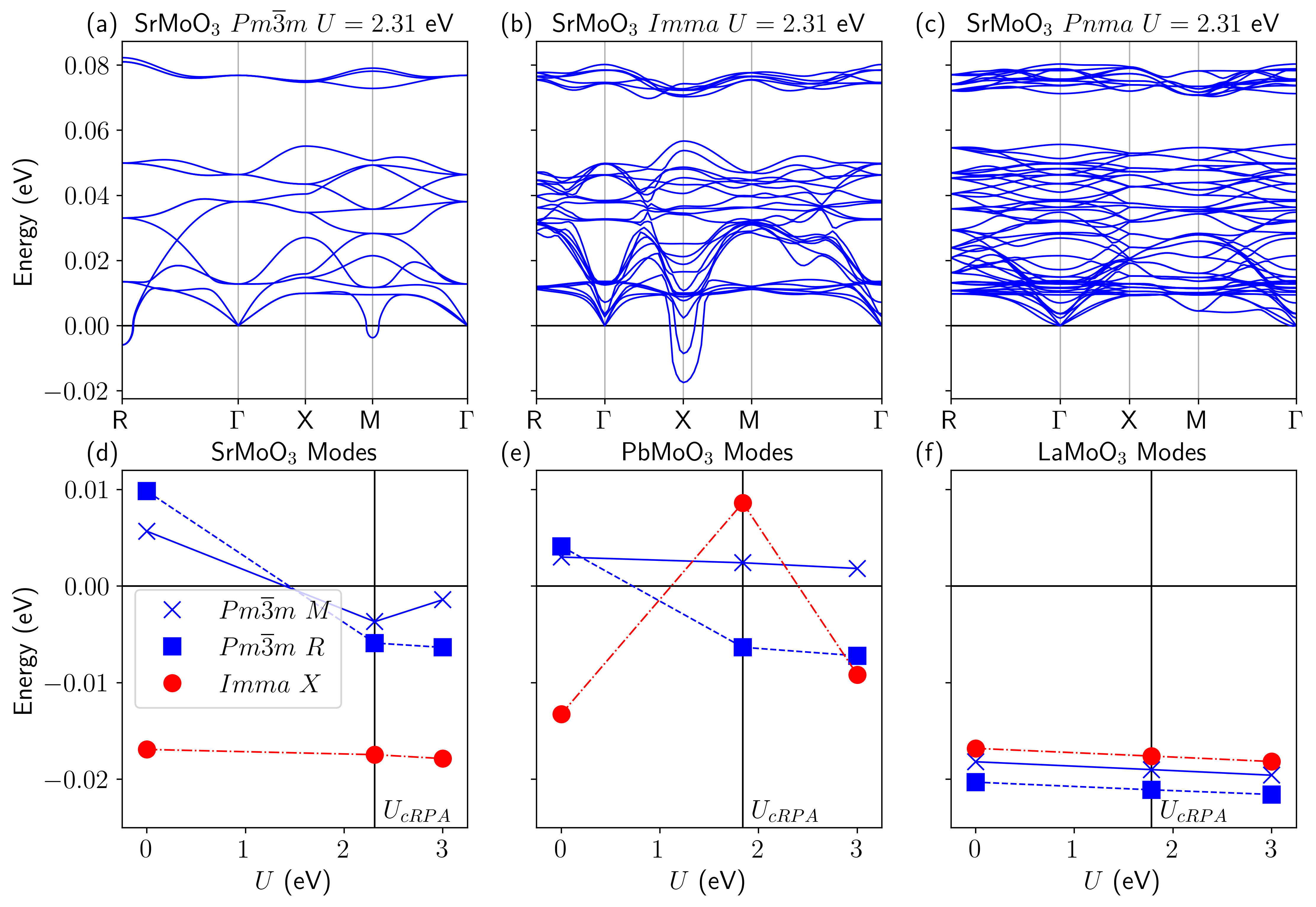}
    \caption{Phonon dispersions of \smo{} in the (a) \cubic{}, (b) $Imma$ and (c) $Pnma$ structures with $U = 2.31$~eV (i.e., \crpau{}). (d), (e) and (f) show the lowest energy modes in \smo{}, \pmo{} and \lmo{} respectively at the high-symmetry $M$ and $R$ points of \cubic{} and the $X$ point of $Imma$. All calculations are nonmagnetic. Negative energy modes indicate dynamic instabilities.}
    \label{fig:phonons}
\end{figure*}

To explore the stability of the cubic structure, we perform calculations of the phonons for SMO, PMO, and LMO (for now, no magnetic ordering is considered). In Fig.~\ref{fig:phonons}(a) we show the full phonon bandstructure for \cubic{} SMO, calculated with \crpau{}. We see that it has unstable modes (i.e., with negative energies) at the $R$ and $M$ points, which correspond to the $R_4^+$ and $M_3^+$ modes that reduce the symmetry to $Imma$ and $Pnma$, respectively \cite{PerezMato:2010ix,Balachandran:2013cg, Miao2014} (see Fig.~\ref{fig:crystal_structures}). Performing the phonon calculation in the SMO $Imma$ structure [Fig.~\ref{fig:phonons}(b)] shows unstable modes at the X point, whereas all of the modes are stable in $Pnma$ [Fig.~\ref{fig:phonons}(c)]. This indicates that $Pnma$ is dynamically stable for SMO at \crpau{}.

The instabilities in the \cubic{} and $Imma$ structures of SMO depend on the choice of $U$. To illustrate this, we plot in Fig.~\ref{fig:phonons}(d) the the energies for the lowest-energy phonon modes at $M$ and $R$ in the \cubic{} structure, and $X$ in $Imma$, versus $U$. With $U\geq$~\crpau{}, there are unstable modes at both $M$ and $R$ [consistent with Fig.~\ref{fig:phonons}(a)]; this, along with the unstable $X$ mode in $Imma$ for all $U$, illustrates that $Pnma$ is stabilized by increasing $U$.  Figure~\ref{fig:phonons}(e) is the corresponding plot for PMO. We see that the behavior of the unstable $R$ mode in PMO is the same as in SMO. However, the modes at $M$ remain stable for $U=0$ to 3 eV. We can also see that at \crpau{}, the $Imma$ $X$ mode is also stable, indicating that $Imma$ is stabilized at that $U$. Finally, we see that all modes considered are unstable for LMO at all values of $U$ [Figure~\ref{fig:phonons}(f)]. We would like to emphasize that these calculations are performed in the NM state; we will determine the ground-state magnetic ordering in Sec.~\ref{sec:mag_struc}, and use this to determine the ground state structure in Sec.~\ref{sec:atom_struct}.

The fact that LMO is unstable in the cubic structure regardless of $U$, while SMO and PMO require a finite $U$ to destabilize the \cubic{}, suggests that the structural properties of SMO and PMO will be more sensitive to the treatment of correlations than LMO. We will see in  Sec.~\ref{sec:atom_struct} that this is the case. Also, the fact that both $R_4^+$ and $M_3^+$ modes are unstable in \cubic{} LMO, even for $U=0$ hints that $Pnma$, which includes both modes, will be especially stable; we will confirm this with total energy calculations in Sec.~\ref{sec:atom_struct}.

\subsection{Magnetic Structure \label{sec:mag_struc}}

To determine the ground state crystal structure, we must first determine the lowest energy magnetic structure. Motivated by the instability of the \cubic{} structure for $U>0$ discussed in the previous section, we will investigate the magnetic structure of both the cubic and orthorhombic phases. All atomic coordinates are relaxed for each magnetic structure and $U$, with the constraint that the space-group symmetry is maintained.

\begin{figure*}[htb]
    \centering
    \begin{subfigure}[b]{0.32\textwidth}
    \centering
        \includegraphics[width=\textwidth]{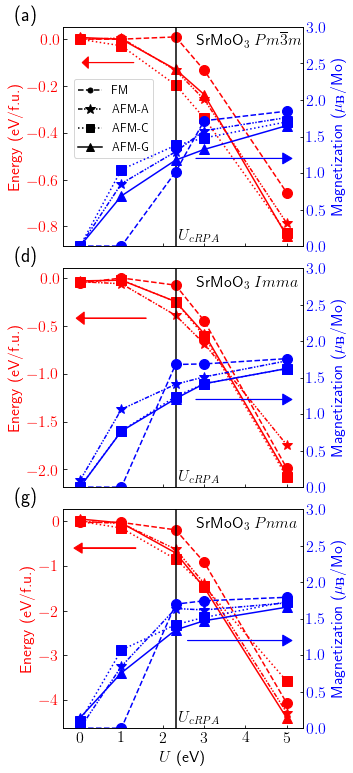}
    \end{subfigure}
        \begin{subfigure}[b]{0.32\textwidth}
    \centering
        \includegraphics[width=\textwidth]{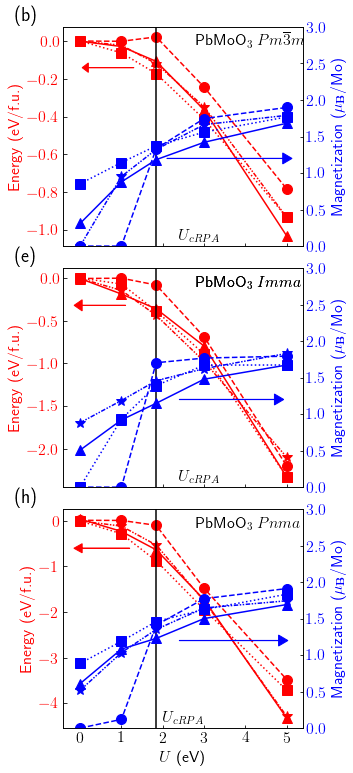}
    \end{subfigure}
        \begin{subfigure}[b]{0.32\textwidth}
    \centering
        \includegraphics[width=\textwidth]{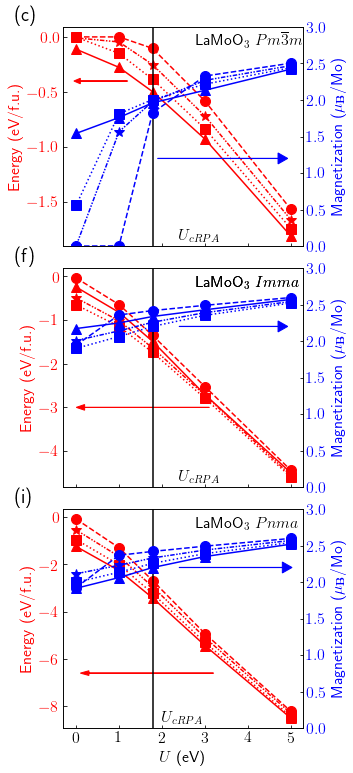}
    \end{subfigure} 
    \caption{Energy of different magnetic orders [ferromagnetic (FM), and antiferromagnetic type-A/type-C/type-G (AFM-A/AFM-C/AFM-G)] referenced to the energy of the nonmagnetic structure and magnitude of the magnetization on the Mo atoms versus Hubbard $U$ for \cubic{} structure of (a) SrMoO$_3$, (b) PbMoO$_3$, and (c) LaMoO$_3$; 
    $Imma$ structure of (d) SrMoO$_3$, (e) PbMoO$_3$, and (f) LaMoO$_3$; $Pnma$ structure of (g) SrMoO$_3$, (h) PbMoO$_3$, and (i) LaMoO$_3$. The solid vertical lines on each plot indicate the $U$ value calculated using the constrained random phase approximation for each material. }  
    \label{fig:en_mag}
\end{figure*}

In Fig.~\ref{fig:en_mag}, we plot the energy of different magnetic orderings (referenced to the energy of the NM state) versus $U$, as well as the magnitude of the magnetic moments on the Mo atoms for \cubic{}, $Imma$, and $Pnma$ structures.

From Fig.~\ref{fig:en_mag}(a), (d), and (g), we see that the qualitative behavior is the same for all structures of SMO. At $U=0$, the Mo magnetic moments vanish, resulting in a NM structure. As we increase $U$, an ordered magnetic state is increasingly favored. Also, the magnetic moment per Mo atom increases with $U$, saturating to a value around 2 $\mu_{\text{B}}$, which is the moment expected for the nominal $4d$ occupation of Mo. At all $U \geq 1$~eV, the AFM state is lower in energy than the FM state, though the specific AFM order is less clear. At \crpau{} we predict AFM-C order for \cubic{} and $Pnma$, and AFM-A order for $Imma$. For both orthorhombic structures, the lowest energy AFM order also changes with $U$, however, even for very large values of $U$ (up to 7 eV), FM ordering is not stabilized.

The picture is similar for PMO [Fig.~\ref{fig:en_mag}(b), (e), and (h)]. Though finite magnetic moments on the Mo may be stabilized even at $U=0$, there is a vanishing energetic benefit to forming an ordered structure. $U$ stabilizes an AFM structure (with Mo moments increasing towards 2 $\mu_{\text{B}}$), also with the exact order being somewhat ambiguous and $U$-dependent. At \crpau{}, the ordering is the same as for SMO: AFM-A order is most stable for $Imma$ PMO, while AFM-C is most stable for $Pnma$ and \cubic{}.

The cubic phase of LMO [Fig.~\ref{fig:en_mag}(c)] is somewhat similar to PMO, where magnetic moments can be stabilized at $U=0$, but with a small energetic driving force for ordering. For $U\geq 1$, the AFM-G state is clearly lower in energy, and the moments increase with $U$ toward, but remain somewhat below 3 $\mu_{\text{B}}$ (expected for the nominal valence of Mo in LMO). The orthorhomic structures of LMO [Fig.~\ref{fig:en_mag}(f) and (i)] are qualitatively different than \cubic{} LMO, and the other materials, in that AFM ordering is more favorable even at $U=0$. For a significant range around \crpau{}, the AFM-C order is the most stable for $Imma$ and the AFM-G order for $Pnma$. The magnetic moment increases slightly with $U$, towards 3 $\mu_{\text{B}}$. 

We can understand the increased stability of magnetic ordering in LMO by considering the electronic structure compared to SMO/PMO. Specifically, the Fermi level in NM LMO is significantly closer to the vHS as a result of the different nominal charge of Mo in LMO ($3+$) versus SMO/PMO ($2+$); $4d-4f$ repulsion may also play a role in LMO. This is indicative of a propensity for instability towards magnetic ordering.

\subsection{Crystal Structure}
\label{sec:atom_struct}

\begin{figure*}[htb]
  \centering
    \begin{subfigure}[b]{0.32\textwidth}
        \centering
        \includegraphics[width=0.95\textwidth]{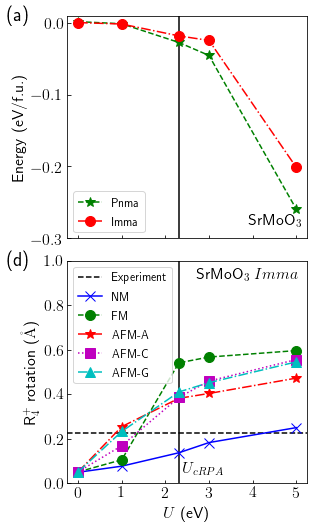}\label{fig:sr_en_rot}
    \end{subfigure}
    \begin{subfigure}[b]{0.32\textwidth}
        \centering
        \includegraphics[width=0.95\textwidth]{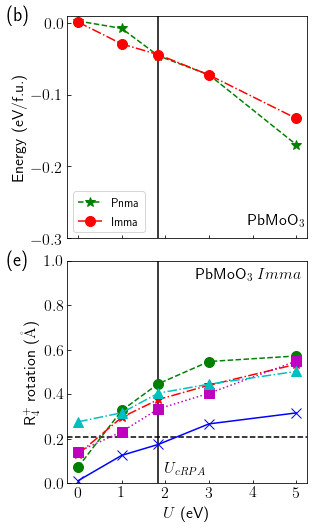}\label{fig:pb_en_rot}
    \end{subfigure}
    \begin{subfigure}[b]{0.315\textwidth}
        \centering
        \includegraphics[width=0.95\textwidth]{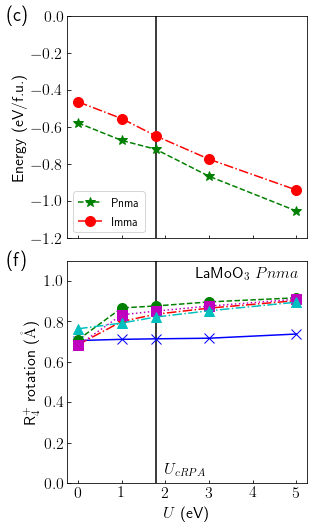}\label{fig:la_en_rot}
    \end{subfigure}
    \caption{Energy of $Imma$ and $Pnma$ structures (in the lowest-energy magnetic state) referenced to \cubic{} for (a) SrMoO$_3$, (b) PbMoO$_3$, and (c) LaMoO$_3$ (not the different scale on the $y$ axis). Amplitude of the rotation mode $R_4^+$ for (d) $Imma$ SrMoO$_3$, (e) $Imma$ PbMoO$_3$, and (f) $Pnma$ PbMoO$_3$. Experimental rotation amplitudes for  (d) SrMoO$_3$ (Ref.~\onlinecite{Macquart2010}), and for (e) PbMoO$_3$ (Ref.~\onlinecite{Takatsu2017}) are shown as dashed horizontal lines. The solid vertical lines on each plot indicate the $U$ value calculated using the constrained random phase approximation for each material.}\label{fig:en}
\end{figure*}

Now that we have determined the lowest-energy magnetic structure for the various crystal structures of SMO, PMO, and LMO, we compare in Fig.~\ref{fig:en}(a)-(c) the energy of $Imma$ and $Pnma$ with respect to $Pm\overline{3}m$ for different choices of $U$ (all in their most favorable magnetic structure). A negative energy indicates that the given orthorhombic structure is more stable than $Pm\overline{3}m$ for that value of $U$. Consistent with the phonon calculations in Sec.~\ref{sec:phonons}, we see that the energy of the orthorhombic structures are lower than cubic for $U\geq1$ for all materials, and also at $U=0$ for LMO. For SMO and PMO, the energies of $Imma$ and $Pnma$ are very close in energy around \crpau{}. We see in Fig.~\ref{fig:en}(a) and (b) that, for SMO at \crpau{}, the $Pnma$ structure is approximately 10 meV/formula unit (f.u.) lower in energy than $Imma$ while for PMO the opposite is true. For LMO [Fig.~\ref{fig:en}(c)], the $Pnma$ structure is more clearly lower in energy than $Imma$ (by $70$ meV/f.u.) at \crpau{}, and this ordering is not sensitive to the choice of $U$.

In Fig.~\ref{fig:en}(d)-(f), we plot the $R_4^+$ oxygen octahedral rotation mode (the most significant amplitude mode that takes the \cubic{} structure to $Imma$), for the different magnetic orderings. There is an additional $R_5^+$ mode allowed by $Imma$ symmetry, which is usually much smaller than the $R_4^+$ amplitude (see Table \ref{tab:modes}). A $M_3^+$ mode and $X_5^+$ mode takes $Imma$ to $Pnma$ ~\cite{PerezMato:2010ix,Balachandran:2013cg, Miao2014} (Table \ref{tab:modes}) which we will also comment on. To compare with experimentally measured rotations for SMO and PMO [horizontal dashed lines in Fig.~\ref{fig:en}(d) and (e)], we give the $R_4^+$ amplitude in the $Imma$ structure (the results for $Pnma$ give the same qualitative picture, e.g., see Fig.~\ref{fig:sr_hse} in Appendix \ref{sec:HSE}), while for LMO, we plot the magnitude for $Pnma$, since it is clearly the most stable structure [see Fig.~\ref{fig:en}(c)]. See Table \ref{tab:modes} for values of the mode amplitudes.

In all of the materials the octahedral rotations are enhanced by $U$, especially when magnetic order is present. For SMO, we find that the structure relaxes to \cubic{} at $U=0$, as the rotation amplitude vanishes [Fig.~\ref{fig:en}(d)], consistent with the dynamical stability shown in Fig.~\ref{fig:phonons}(d). This is the case regardless of magnetic structure, since, as we saw in Fig.~\ref{fig:en_mag}(a), the Mo magnetic moments vanish at $U=0$, and thus all magnetic structures converge to NM. At \crpau{}, the $R_4^+$ amplitude for SMO in its lowest-energy (AFM-A) magnetic ordering is significantly overestimated compared to experiment \cite{Macquart2010} (see Sec.~\ref{sec:discuss} for detailed comparison with experiment). The rotations in the NM state are closer to experiment at around \crpau{}, but we have shown in Fig.~\ref{fig:en_mag}(a) and (d), that the NM structure is highly energetically unfavorable at this $U$. The overestimation of the octahedral rotations is also not specific to our treatment of the Coulomb interaction with PBE+$U$. We show in Appendix \ref{sec:HSE} that we see similar behavior using the HSE hybrid functional, i.e., the AFM ordering is lowest in energy, and the $R_4^+$ amplitude is $0.4-0.5$ \AA{}.

For PMO [see Fig.~\ref{fig:en}(e)], the structure also tends toward cubic at $U=0$, though small rotations ($\sim 0.1$ \AA{}) persist. These rotations are larger for the magnetically ordered states. As with SMO, the octahedral rotations increase with $U$, and the amplitude predicted by PBE+\crpau{} is significantly larger than the experimental \cite{Takatsu2017} one for the lowest-energy AFM-A magnetic structure. Again, the NM calculation is more in line with the experimental $R_4^+$ amplitude, but is energetically unfavorable [Fig.~\ref{fig:en_mag}(b)].

As we expect from the stability of the $Pnma$ structure for LMO shown in Fig.~\ref{fig:en}(c), the octahedral tilts are present and large for LMO at all values of $U$ [Fig.~\ref{fig:en}(f)].

 We can make a quantitative comparison between the $Pnma$ structure of the molybdates by comparing their $M_3^+$ and $X_5^+$ modes, which differentiate $Imma$ and $Pnma$, as mentioned above (see Table \ref{tab:modes}). Not surprisingly, LMO has very significant $M_3^+$ and $X_5^+$ amplitudes that do not change much between $U=1$ and $U=3$ eV, reflecting the stability of the $Pnma$ structure of this material. PMO, however, has very small amplitudes of $M_3^+$ and $X_5^+$, indicating that the $Pnma$ structure has only slight deviations from $Imma$. This is consistant with the very small energy difference between $Imma$ and $Pnma$ for PMO seen in Fig.~\ref{fig:en}(b). Finally, the $Pnma$-related distortion amplitudes for SMO are intermediate between LMO and PMO. They are also more sensitive to $U$, increasing significantly between $U=1$ and $U=3$ eV. This analysis clearly indicates that the mode softening with $U$ in the cubic structure of SMO and PMO results in a nontrivial dependence of the structural properties on the correlations. Since the magnetic properties also depend sensitively on $U$, there is further coupling between the magnetic and structural aspects of these materials. For LMO in the orthorhombic structures, this coupling is much less significant, as magnetic ordering is stabilized by the proximity of the vHS to the Fermi level, and thus the structural properties do not show the same sensitivity to correlation.

\begin{figure}
    \centering
    \includegraphics[width=\linewidth]{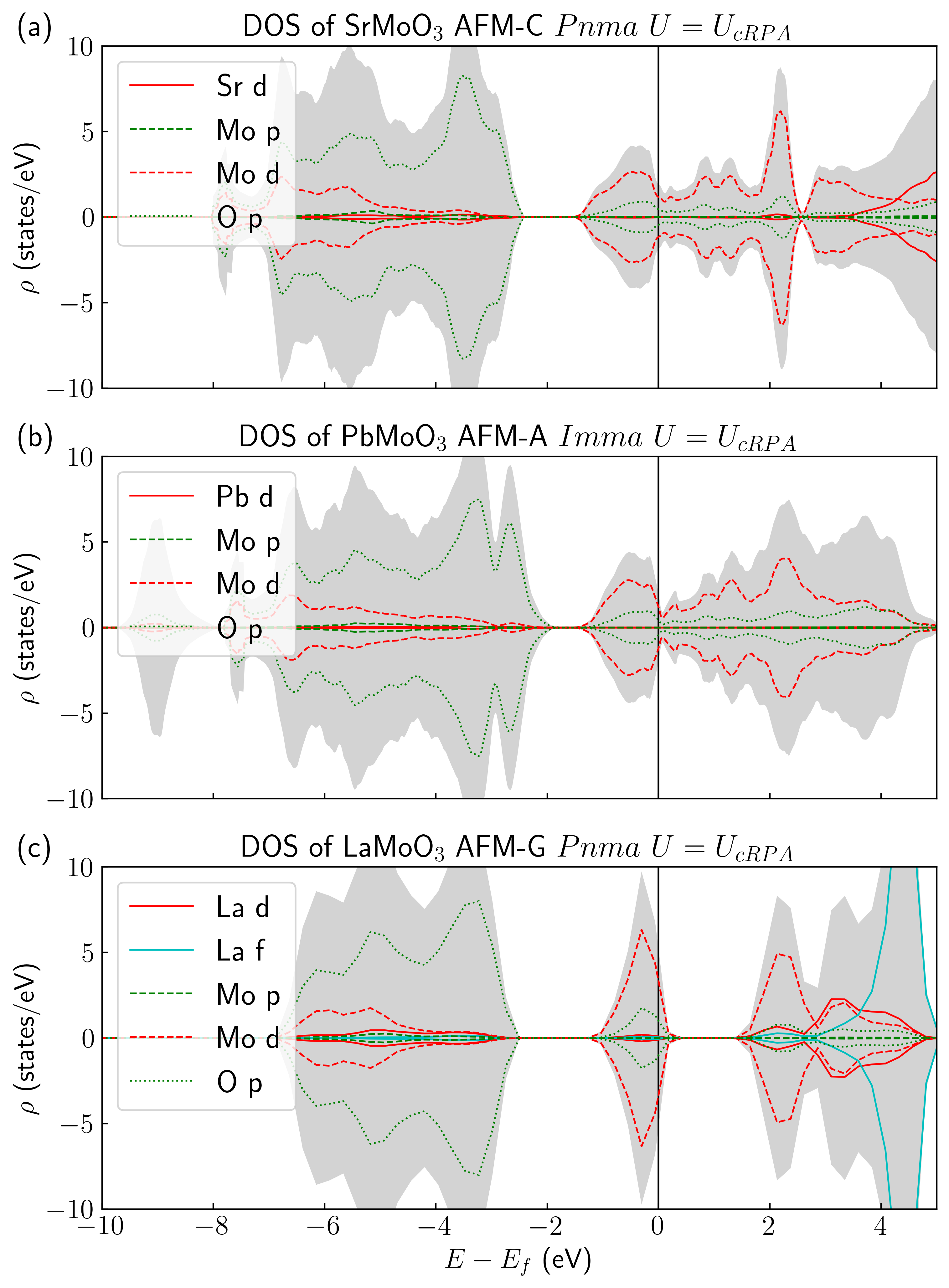}
    \centering
    \caption{Density of states of \smo{}, \pmo{} and \lmo{} in their lowest-energy crystal and magnetic structures at $U_{cRPA}$, i.e., (a) \smo{} in the $Pnma$ AFM-C, (b) \pmo{} in the $Imma$ AFM-A, and (c) \lmo{} in the $Pnma$ AFM-G states.}
    \label{fig:lowest_en_dos}
\end{figure}

Finally, we can comment on the electronic structure of the molybdates in their predicted ground-state crystal/magnetic structure compared to the cubic cases discussed in Sec.~\ref{sec:elec_struct} (see Fig.~\ref{fig:lowest_en_dos}). As for the cubic case (Fig.~\ref{fig:band}), the states at the Fermi level have Mo $4d$ character, with a gap to the lower-energy O $2p$ states. For LMO, a significant gap also opens between the Mo $4d$ $t_{2g}$ and $e_g$ states, but the Fermi level remains in the lower manifold of states, for a wide range of interaction parameters. Indeed, the metallic character of all of the molybdates remains for these orthorhombic structures, and is very robust against a wide range of choices of $U$ and $J$ in the calculations.

\begin{table*}[htpb]
\begin{ruledtabular}
\caption{Amplitudes of distortion modes from the cubic to orthorhmobic structures for different values of $U=1$~eV, \crpau{}, and $U=3$~eV, and experimentally measured amplitudes from Ref.~\onlinecite{Macquart2010} and Ref.~\onlinecite{Takatsu2017}. For calculations, the predicted magnetic ordering is specified.}
    \label{tab:modes}
    \begin{tabular*}{\textwidth}{c|c|c|c|cccc}
        & Struct. & Mag.~Ord. & $U$ & $R_4^+$ & $M_3^+$ & $X_5^+$ & $R_5^+$  \\
        \hline
         \multirow{4}{*}{SrMoO$_3$} & \multirow{2}{*}{$Imma$} & \multirow{3}{*}{AFM-A} & 1 & 0.254 & -- & -- &  0.006 \\ & & & 2.31 & 0.346 & -- & -- &  0.016 \\ & & & 3 & 0.403 & -- & --  & 0.040 \\\cline{3-8}
        & & -- & Exp.\footnote{Ref.~\onlinecite{Macquart2010}} & 0.226 & -- & --  & 0.012 \\
         \cline{2-8}
        & \multirow{2}{*}{$Pnma$} &  \multirow{3}{*}{AFM-C} & 1 & 0.218 & 0.140 & 0.029 & 0.003\\
        & & & 2.31 & 0.265 & 0.252 & 0.045 & 0.003 \\
        & & & 3 & 0.346 & 0.331 & 0.112 & 0.020 \\
         \hline
         \multirow{4}{*}{PbMoO$_3$}& \multirow{2}{*}{$Imma$ } & \multirow{3}{*}{AFM-A} & 1 & 0.395 & -- & -- & 0.110 \\ 
         & & & 1.84 & 0.439 & -- & --  & 0.144\\
         & & & 3 & 0.475 & -- & --  & 0.071\\\cline{3-8} & & --  & Exp.\footnote{Ref.~\onlinecite{Takatsu2017}} & 0.207 & -- & --  & 0.004 \\
         \cline{2-8}
         & \multirow{2}{*}{$Pnma$} &  \multirow{3}{*}{AFM-C}& 1 & 0.299 & 0.05 & 0.011 &  0.050 \\ 
         & & & 1.84 & 0.434  & 0.025 & 0.007 &  0.143\\ 
          & & & 3 & 0.477 & 0.043 & 0.012 & 0.169 \\
         \hline
         \multirow{4}{*}{LaMoO$_3$} & \multirow{2}{*}{$Imma$} & \multirow{3}{*}{AFM-C} & 1 & 0.890 & -- & -- & 0.120 \\
         & & & 1.78 & 0.947 & -- & -- & 0.139 \\
         & & & 3 & 0.964 & -- & -- & 0.138 \\
         \cline{2-8}
         & \multirow{2}{*}{$Pnma$ } & \multirow{3}{*}{AFM-G} & 1 & 0.777 & 0.552 & 0.376 & 0.079 \\
         & & & 1.78 & 0.802 & 0.558 & 0.380 & 0.081 \\
         & & & 3 & 0.835 & 0.565 & 0.384 & 0.084 \\
    \end{tabular*}
    \end{ruledtabular}
\end{table*}

\section{Discussion}
\label{sec:discuss}

To summarize the findings in Sec.~\ref{sec:results}, our PBE+$U$ calculations predict a metallic electronic structure, an orthorhombic space group, and AFM ordering for all SMO, PMO, and LMO. It is clear from Figs.~\ref{fig:en_mag} and \ref{fig:en} that the structure, magnetic ordering, and treatment of the Coulomb interaction are coupled to each other. This is especially relevant in SMO and PMO, where the sensitivity of the structural (see also Table \ref{tab:modes}) and magnetic properties with Hubbard $U$ is much more significant than for LMO. Increasing $U$ results in a simultaneous increase in the stability of the magnetic order and magnitude of the magnetic moments (Fig.~\ref{fig:en_mag}), as well as the stability of the orthorhombic structure and magnitude of the distortions (Fig.~\ref{fig:en}). In this section, we will discuss the interpretation of these results in the context of experimental observations and similar perovskite oxides.

\subsection{Tolerance factor \label{sec:tol}}
An empirical approach for predicting crystal structures of perovskites is based on the ``tolerance factor'' $t$, which predicts the ratio of lattice constants by considering the ionic radii of the constituents:
    \begin{equation}
    \label{eq:tol}
        t = \frac{(r_A + r_{\text{O}})}{\sqrt{2} (r_A + r_{B})},
    \end{equation}
where $r_x$ is the ionic radii for the atom $x$. In a hard-sphere model, a tolerance factor close to unity would indicate a stable cubic $Pm\overline{3}m$ ionic packing. Values of $t < 1$ predict a distortion away from the cubic structure towards an orthorhombic structure~\cite{Goldschmidt1926}.

Taking ionic radii from Ref.~\onlinecite{Shannon1976}, we give values for the tolerance factor in the second column of Table \ref{tab:tolerance}. The range of values for SMO, PMO and LMO do not deviate much from the cubic structure in the ideal ionic picture. Thus the structural distortions are not explained simply by the ionic hard-sphere model.

We also report in the third and fourth columns of Table \ref{tab:tolerance} the tolerance factors calculated using bond lengths from the lowest energy structures of the molybdates at $U = 0$ and \crpau{}. We see that for SMO and PMO, increasing the Coulomb interaction shrinks the effective radii of the Mo atom, decreasing $t$ and favoring the orthorhombic structure. For LMO, $U$ does not have a significant influence on the relative bond lengths.

\begin{table}[htpb]
\begin{ruledtabular}
\caption{Tolerance factor $t$ [see Eq.~(\ref{eq:tol})] of SrMoO$_3$, PbMoO$_3$, and LaMoO$_3$ using data from ~\citet{Shannon1976} (Ionic radii column) and the lattice parameters obtained in this study for $U=0$~eV and $U=$ \crpau{}. A value of $t \simeq 1$ predicts a stable \cubic{} structure. }
    \label{tab:tolerance}
    \begin{tabular}{c|ccc}
        & $t$ (Ionic radii)  & $t$ ($U = 0$) &  $t$ ($U = U_{\text{cRPA}}$) \\
        \hline
         SrMoO$_3$ &  0.979 & 0.999  & 0.929 \\ 
         PbMoO$_3$ &  0.996 & 0.973 & 0.926 \\ 
         LaMoO$_3$ &  0.952 & 0.819 & 0.817\\ 
    \end{tabular}
    \end{ruledtabular}
\end{table}

\subsection{Comparison with experimental observations}

As we summarized in Sec.~\ref{sec:exp}, SMO and PMO are found experimentally to be metallic~\cite{Brixner1960,Nagai2005,Takatsu2017} with no magnetic ordering~\cite{Ikeda2000,Macquart2010,Zhao2017} detected (even at quite low temperatures), and found to be in the $Imma$ crystal structure at low $T$~\cite{Macquart2010, Takatsu2017}. 

We indeed find all of the molybdates studied here to be metals, which is robust to the choice of the $U$ parameter. We also predict orthorhombic structures, though we cannot clearly discern between $Imma$ and $Pnma$ (the latter is not detected in experiment) for SMO and PMO. Structurally (see Table \ref{tab:modes}), $Pnma$ and $Imma$ for PMO are almost identical (measured in $M_3^+$ and $X_5^+$ mode amplitudes), while the structural distinction is sensitive to the choice of $U$ for SMO. We clearly predict AFM ordering for all of the materials at reasonable values of $U$ around \crpau{} (see Fig.~\ref{fig:en_mag}), whereas in experiment a paramagnetic state is observed. In addition, our calculations overestimate the $R_4^+$ rotation amplitude by about a factor of two compared to experiment \cite{Macquart2010,Takatsu2017} [Fig.~\ref{fig:en}(a) and (b)]. As discussed in Sec.~\ref{sec:atom_struct}, artificially suppressing the magnetization in our calculations [blue solid curve in Fig.~\ref{fig:en}(a) and (b)] results in an $R_4^+$ amplitude closer to experiment, but the non-magnetic calculation is significantly energetically unfavorable. 

As we discussed in the beginning of Sec.~\ref{sec:discuss}, we have demonstrated the close coupling in SMO and PMO between the Coulomb interaction, atomic, and magnetic structure. Thus, the reason for the discrepancy in the structural and magnetic properties between experiment and theory may be due to the fact that we cannot directly model a \emph{paramagnetic} state, which was what is measured in SMO and PMO \cite{Macquart2010,Ikeda2000,Zhao2017}. This would require, e.g., a dynamical mean-field theory (DMFT) treatment \cite{ABRIKOSOV201685, RevModPhys.68.13,Wadati2014,Paul2019,Hampel2020}. In principle, a paramagnetic state can be approximated by averaging over ensembles of spin configurations \cite{Varignon2019}, however obtaining structural properties from such calculations would be a significant challenge, as well as reproducing the Pauli paramagnetic state observed in experiment. Because of the sensitivity of the structural properties of SMO and PMO, neither the magnetically ordered states, nor the NM case are a good enough approximation to the experimental electronic structure to produce an accurate prediction of the structure. Similar effects have been observed in rare-earth nickelates for the magnitude of the breathing mode distortion comparing DFT+$U$ and DFT+DMFT~\cite{Park2014short, Haule2017,hampel:2019}.

In the context of the coupling between the magnetic and atomic structure, it is important to understand the origin of the stabilization of the magnetic order in DFT+$U$. We have shown in Sec.~\ref{sec:mag_struc} that at least for SMO and PMO, $U$ drives the stabilization of the magnetic moments [in LMO, significant moments occur at $U=0$ in many cases, see Fig.~\ref{fig:en_mag}(c), (f), and (i)]. Since the DFT+$U$ calculations are conducted self consistently, the exchange splitting caused by the Hubbard $U$ may be enhanced by the DFT exchange-correlation potential. To determine the role of this effect, we can turn off the spin polarization in the DFT part of the calculation, so all of the exchange splitting is from the Hubbard $U$. If we do this for SMO (for \crpau{}), we indeed find smaller octohedral rotations (0.273 \AA{}), more similar to the experimental value (0.226 \AA{}~\cite{Macquart2010}). However, the ground-state magnetic order remains AFM-C in the $Pnma$ space-group.

Finally, we comment on the fact that LMO has yet to be successfully synthesized in the perovskite structure/stoichiometry. We show in Fig.~\ref{fig:stability_diagram} in Appendix \ref{sec:lmo_stab} that LMO is actually not thermodynamically stable. For Mo and O rich conditions, LMO is unstable toward the formation of MoO$_2$, while for Mo and O poor conditions, it is unstable towards the formation of La$_2$O$_3$. Among the intermediate chemical potentials, we find a region where La$_2$Mo$_3$O$_{12}$ is stable, which was the stoichiometry synthesized in Ref.~\onlinecite{Figen2014}. Even though LMO is not stable in bulk form, it may be stabilized as a thin film grown on a perovskite substrate, or in a superlattice with other perovskites. In addition, La may be alloyed with SMO or PMO, and our findings (i.e., Fig.~\ref{fig:en_mag}) indicate that such alloying may stabilize magnetic ordering in SMO or PMO, tune the temperatures of the structural transitions by stabilizing the orthorhombic phases, or otherwise result in interesting correlation effects by shifting the Fermi level closer to the vHS~\cite{Karp2020} as it is in LMO.

\subsection{Comparison with ruthenate and chromate perovskites}

To elucidate the effects of $d$-site occupation and $p-d$ splitting (i.e., the charge-transfer gap), we compare with previous studies on perovskite oxides with Ru or Cr on the $B$ site. In general, the ruthenate and chromate perovskites have a similar electronic structure to their molybdate counterparts, i.e., the states near the Fermi level are predominately hybridized O $2p$ and $B$-site $3d/4d$ $t_{2g}$ states \cite{Kimber2009, Allen1996, Lee2009, Wang2010}, and most calculations \cite{Rondinelli2008,Lee2009, Wang2010, Dabaghmanesh2017} and experiments \cite{Callaghan1966,Allen1996,Chamberland1972,Fujioka1887,Kimber2009} find these materials to be metals (with some notable exceptions described below).

SrRuO$_3$ crystallizes in the the $Pnma$ structure \cite{Jones1989}, as we find to be lowest energy for SMO around \crpau{}. However, SrRuO$_3$ exhibits ferromagnetic order in experiment \cite{Callaghan1966,Kanbayasi1976,Joy1998,Rondinelli2008} and in DFT even without the application of any Hubbard $U$ \cite{Kimber2009,Allen1996,Fujioka1887}. This is likely driven by the fact that the Fermi level is in close proximity to the $4d$ vHS \cite{Fujioka1887}, unlike in SMO, where it is 1~eV below the vHS [Fig.~\ref{fig:band}(a)]. SrCrO$_3$ is found to be either a cubic paramagnet \cite{Chamberland1967} (down to 4.2 K), a tetragonal antiferromagnet \cite{Zhou2006,Ortega-San-Martin2007} (with a transition temperature around $35-40$~K  \cite{Ortega-San-Martin2007}), or a coexisistance of the two  \cite{Ortega-San-Martin2007,Lee2009}. The lack of an orthorhombic phase is consistent with the its tolerance factor being $t > 1$ (versus $t = 0.979$ for SMO, Sec.~\ref{sec:tol}), due to the more tightly bound Cr $3d$ orbitals. As with SMO, the Fermi level is significantly far away from the vHS (0.5 eV above in the cubic NM structure~\cite{Lee2009}), which is likely why it does not exhibit ferromagnetism like SrRuO$_3$ does.

PbRuO$_3$ exhibits an orthorhombic structure like PMO, however, with a temperature-induced transition from metallic $Pnma$ to insulating $Imma$~\cite{Kimber2009}. Similarly to SrRuO$_3$, the vHS in PbRuO$_3$ is very close to the Fermi level for the $Pnma$ structure, and DFT + $U$ calculations show FM and antiferromagetic type-G (AFM-G) solutions are similar in energy, with the AFM-G energy found to be slightly lower\cite{Kimber2009}. Similarly to PMO experiments, no magnetization has been measured in this material down to 1.5 K~\cite{Kimber2009}. PbCrO$_3$ has also been shown to form an AFM-G type order when synthesized in the cubic structure~\cite{Roth1967}, with resistivity indicating semiconducting behaviour~\cite{Chamberland1972}. LDA+$U$ and GGA+$U$ studies with $U = 4$~eV reproduce the antiferromagnetic ground state, but find PbCrO$_3$ to be metallic \cite{Wang2010}. 

Experimental synthesis of LaCrO$_3$ show a $Pnma$ structure below 530~K with AFM-G order below 288 K \cite{Oikawa2000, Zhou2011,Tiwari2015}; this is similar to the ground state we predict for LMO, however resistivity measurements in thin-films indicate a semiconductor with a band gap of approximately $2.8$ eV~\cite{Sushko2013} (consistent with DFT+U calculations~\cite{Dabaghmanesh2017} with $U = 3.3$~eV). A possible reason for this is, unlike LMO, the Lanthanum $f-$states are further away from the Fermi-level and are not considered in calculations. The $f-d$ repulsion in LMO may serve to stabilize the metallic state.

Comparison to these results reveals more homogenous behaviour in the different molybdates than their ruthenate and chromate counterparts. The properties that we predict, particularly in SMO and PMO are similar in contrast to the pairs of SrRuO$_3$/PbRuO$_3$ and SrCrO$_3$/PbCrO$_3$ whose properties vary in experimental and computational results. 

\section{Conclusions}
\label{sec:conclusion}

We have performed systematic calculations of the intertwinded electronic, magnetic, and crystal structures of perovskites SrMoO$_3$, PbMoO$_3$, and LaMoO$_3$ using PBE+$U$. Regardless of the choice of $U$, we find that all three materials are metallic. For $U\simeq U_{\text{cRPA}}$, all materials are predicted to be antiferromagnetic and in an orthorhombic phase: $Pnma$ for SMO and LMO, and $Imma$ for PMO. The $R_4^+$ octahedral rotations are calculated to be significantly larger than observed in experiment for SMO and PMO. The fact that the orthorhombic structure is predicted for all three materials even though the tolerance factor is close to unity indicates that there are non-ionic interactions which are important in these materials.

We find that Coulomb interaction significantly influences the magnetic structure; the stability of the AFM order, as well as the Mo magnetic moments increase with $U$. The magnetic order and Coulomb interaction also influences the structural properties, as the stability of the orthorhombic structures and $R_4^+$ octahedral rotations also increase with $U$. The qualitative effects of the Coulomb interaction are smaller for LMO compared to the other two materials, as it is found to be AFM and $Pnma$ regardless of $U$.

The difference between our findings for SMO and PMO and the experimental observations could be attributed to the lack of dynamic correlations, which we would expect to stabilize a paramagnetic structure. This is left open for future investigations. 

\acknowledgements

We thank K. Rabe for insightful discussions and B. Monserrat for discussions and for sharing a copy of the phonon code {\sc Caesar}. CED and JLH acknowledge support from the National Science Foundation under Grant No. DMR-1918455. The Flatiron Institute is a division of the Simons Foundation. We also thank Stony Brook Research Computing and Cyberinfrastructure, and the Institute for Advanced Computational Science at Stony Brook University for access to the high-performance SeaWulf computing system, which
was made possible by a National Science Foundation grant No. 1531492.

\appendix

\section{Influence of lanthanum  $4f$ states on LaMoO$_3$}
\label{sec:la_f_states}

\begin{figure}
    \centering
    \includegraphics[width=\linewidth]{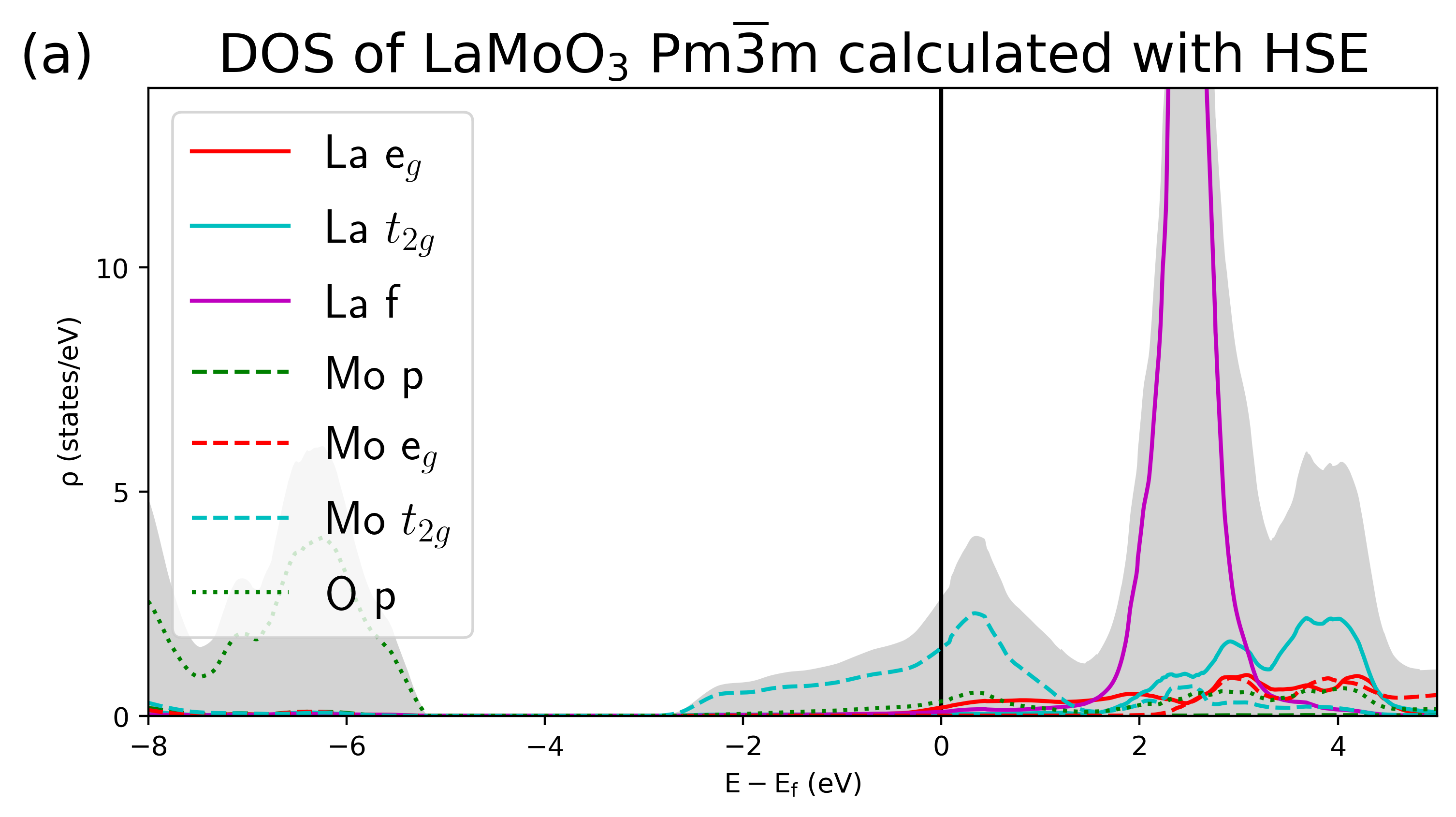}
    \centering
    \includegraphics[width=\linewidth]{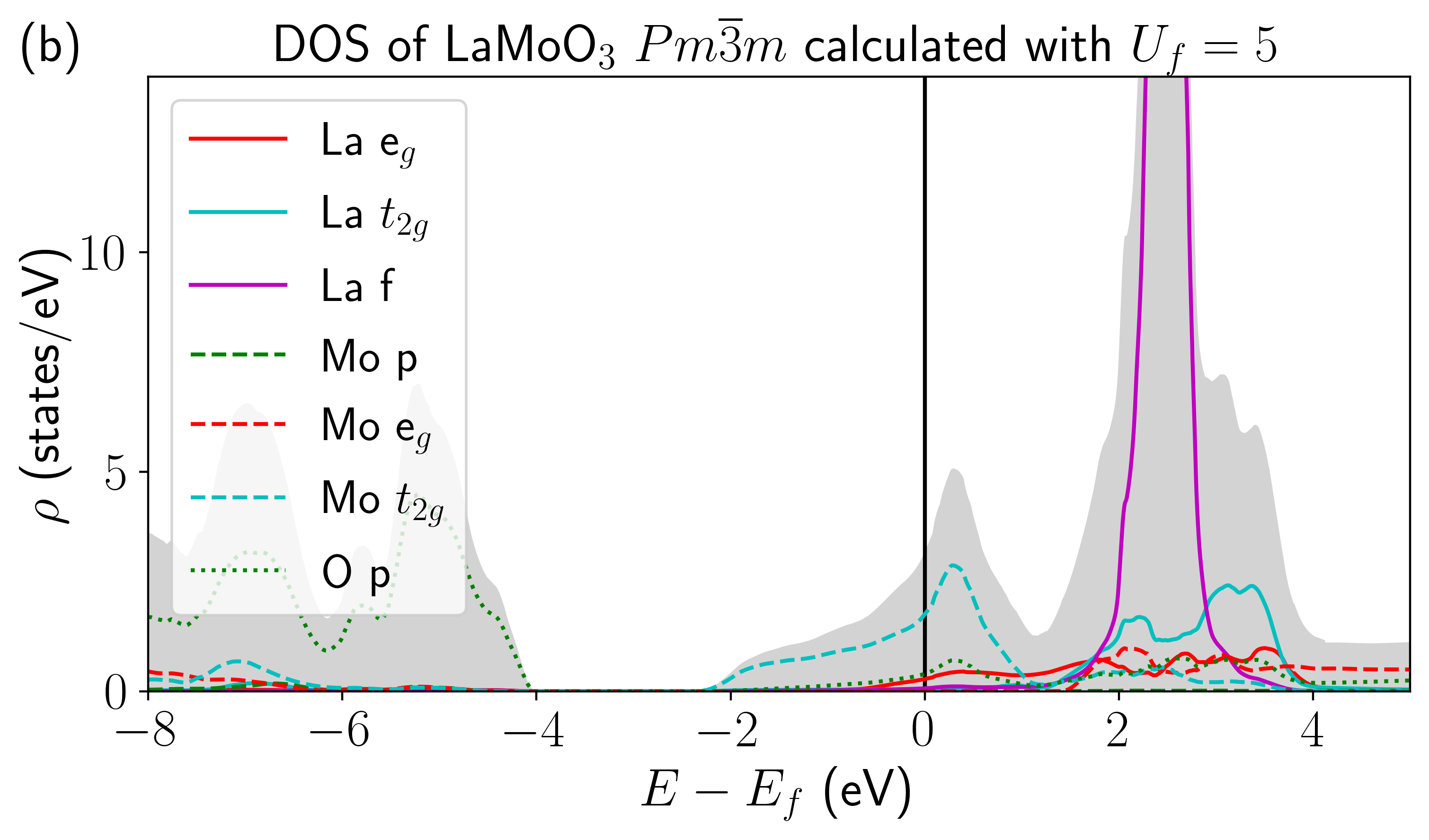}
    \centering
    \includegraphics[width=\linewidth]{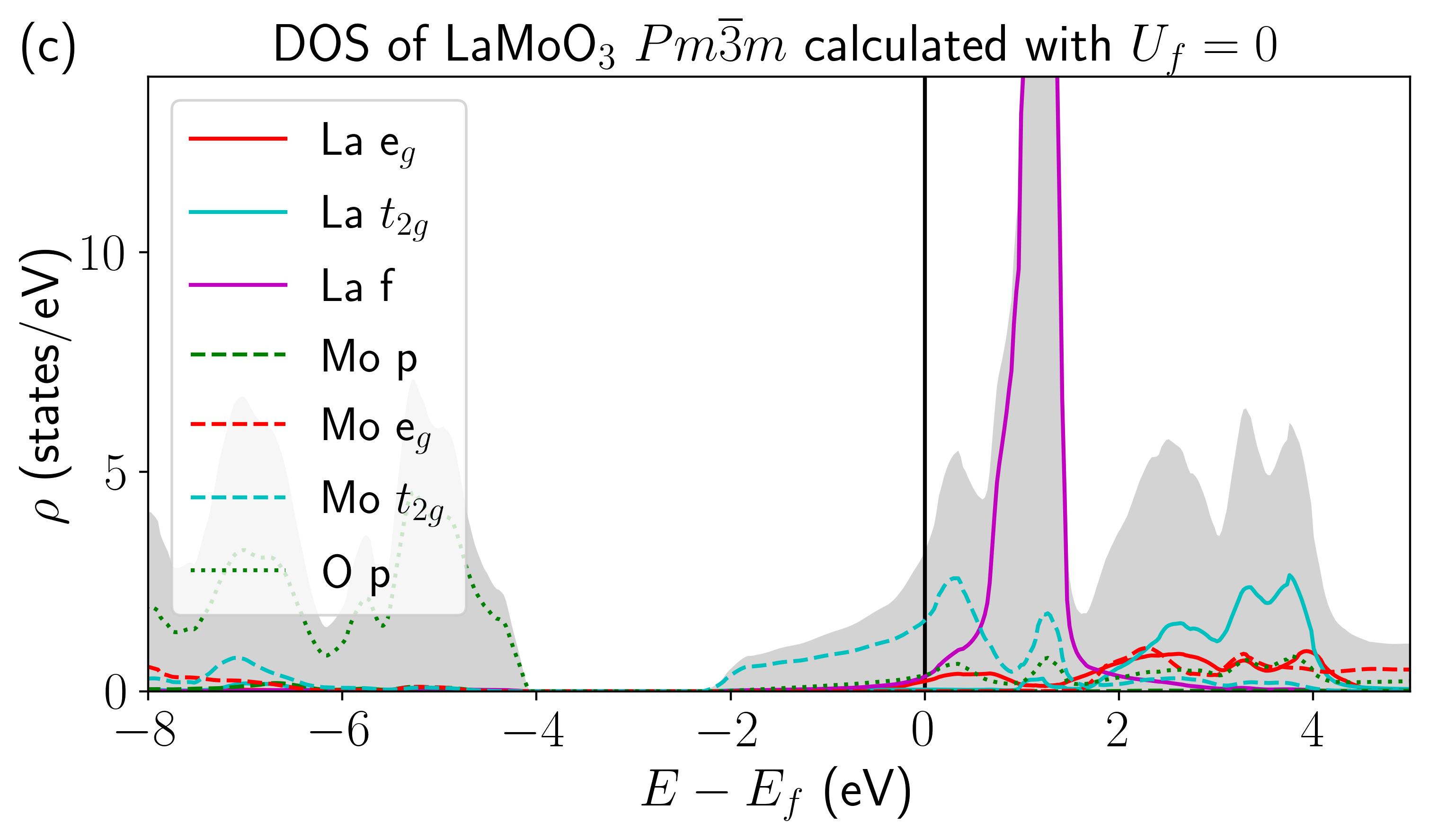}
    \caption{Density of states of LaMoO$_3$ $Pm\overline{3}m$ calculated using (a) HSE, and PBE + $U_f$ with (b) $U_f = 5$~eV and (c) $U_f = 0$~eV on the La $4f$ states.}
    \label{fig:la_f_dos}
\end{figure}

When no $U$ is applied, the empty La $4f$ states are quite close ($<2$ eV) to the Fermi level, which is one of the major differences in the electronic structure of LMO and SMO/PMO. It is likely that the properties of these states, including their relative energy placement is not correctly captured in GGA. Therefore, we explore the effects of placing a Hubbard $U$ also on these orbitals. In order to choose an appropriate value of $U_f$, we compare to the electronic structure of LMO calculated with the HSE \cite{Heyd2006} hybrid functional (with the standard $\alpha=0.25$). HSE is expected to improve the description of the localized $d$ and $f$ states over GGA, without having to choose adjustable parameters for each. Fig. \ref{fig:la_f_dos}(a) shows the DOS of LMO $Pm\overline{3}m$ calculated with HSE. We find that a PBE+$U_f$ with $U_f=5$ eV on the La $4f$ states [Fig. \ref{fig:la_f_dos}(b)] shows good qualitative agreement with HSE, compared to no U on the $f$ states [Fig. \ref{fig:la_f_dos}(c)]. Thus we will use this value of $U_f=5$ eV for the LMO calculations in Sec.~\ref{sec:results}. Comparison between Fig. \ref{fig:la_f_dos}(b) and (c) illustrates that the Mo $t_{2g}$ states around the Fermi level are not significantly affected by $U_f$.

\begin{figure}[htp]
    \centering
    \includegraphics[width=0.95\linewidth]{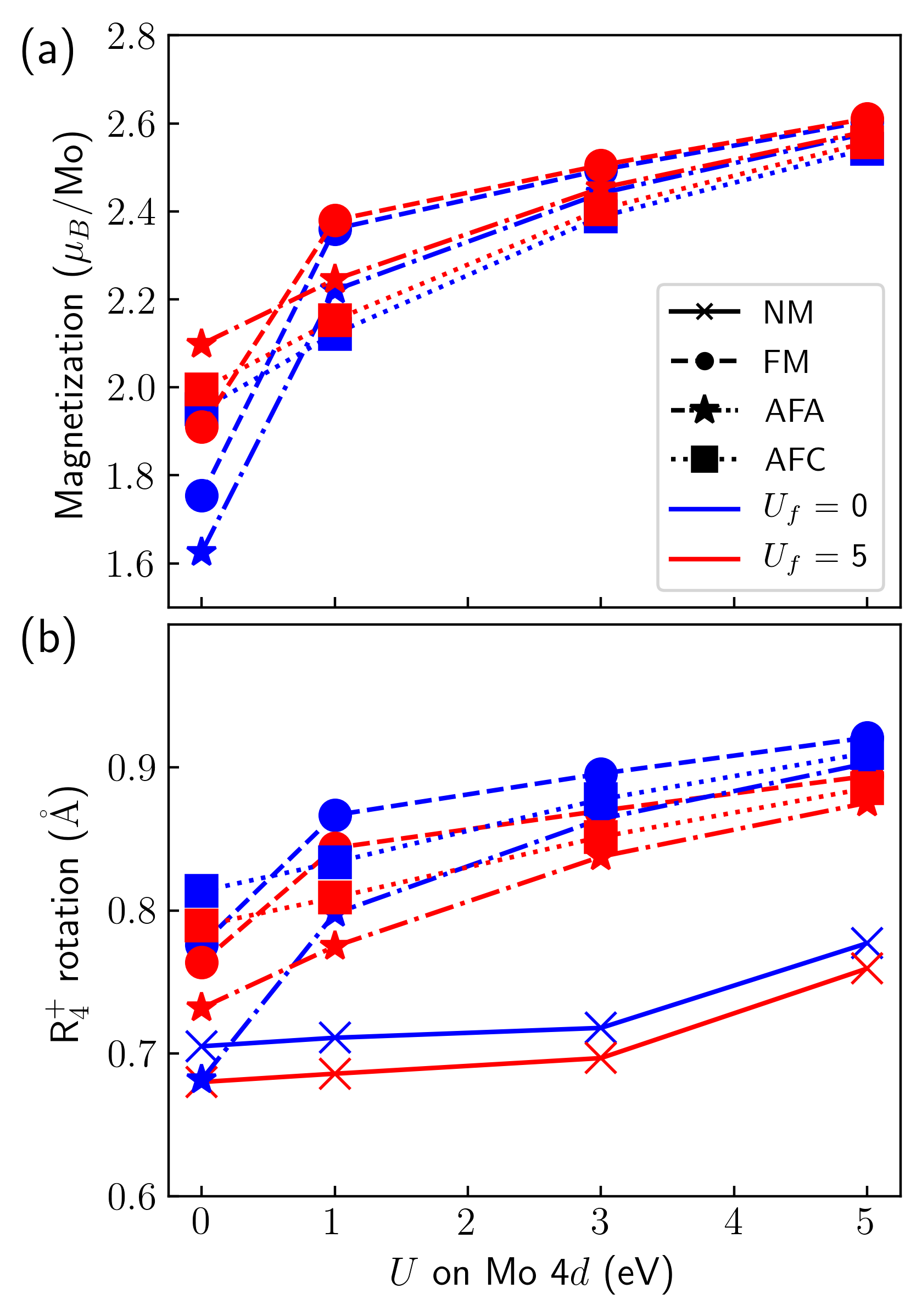}
    \caption{(a) Magnetization of the Mo atoms and (b) amplitude of $R^-_{4}$ octahetral rotation mode for LaMoO$_3$ $Pnma$ with $U_f=0$ and $U_f=5$ eV applied to the lanthanum $f$ states and varying $U$ on the Mo $4d$ states for the various magnetic states.}
    \label{fig:la_f_mag_rot}
\end{figure}

We will now confirm that placing a $U_f$ on the La $4f$ states does not qualitatively effect our conclusions of the magnetic and crystal structure of LMO. Fig.~\ref{fig:la_f_mag_rot}(a) shows the magnetization on the Mo atoms for the $Pnma$ structure versus $U$ on the Mo $4d$ states for both $U_f=0$, and $U_f=5$ eV on the La $4f$ states. We can see that for $U=0$ on the Mo, there is a significant increase in the magnetic moments with the application of $U_f$. However, around \crpau{}, the difference in negligible. 

In Fig.~\ref{fig:la_f_mag_rot}(b), we compare the $R_4^+$ octahedral rotations for $U_f=0$ and $U_f=5$ eV for the different magnetic states. Overall, the application of $U_f$ decreases the rotation amplitudes slightly for all magnetic states. This is at least partially explained by a larger cell volume (of approximately 1\%) in all of these calculations. However, the trends are not changed with or without the application of $U_f$.

\section{HSE Calculations for octahedral rotations in SrMoO$_3$ \label{sec:HSE}}

\begin{figure}[htb]
        \centering
        \includegraphics[width=\linewidth]{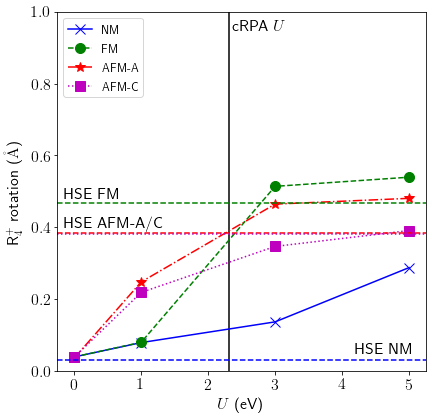}
        \caption{Octohedral tilts in SrMoO$_3$ $Pnma$ PBE + $U$ compared with HSE calculations. Rotations of the octahedra for HSE are shown by horizontal lines. The vertical line indicates the calculated cRPA $U$ value. Note, the amplitude of the rotations of the antiferromagnetic type-A and type-C are nearly degenerate for HSE.}\label{fig:sr_hse}
\end{figure}

To determine the sensitivity of our conclusions to the treatment of the electronic correlations, we compare our PBE+$U$ calculations to those calculated with the HSE hybrid functional~\cite{Heyd2006} (with the default mixing of $\alpha=0.25$). HSE also finds an AFM magnetic state to be lowest in energy. We plot the $R_4^+$ octahedral rotation amplitudes for the $Pnma$ structure in Fig.~\ref{fig:sr_hse}. HSE results are the horizontal dashed lines; note that the amplitude of the AFM-A/C rotation magnitudes calculated with HSE are nearly degenerate. We see that HSE similarly overestimates the $R_4^+$ amplitude compared to experiment as our PBE+\crpau{} calculations. The only significant qualitative difference between HSE and PBE+\crpau{} is that HSE finds negligible octahedral rotations for the NM state, whereas the rotation amplitude is clearly nonnegligible for PBE+\crpau{}. The qualitative agreement between the methods gives us confidence that, e.g., the overestimation of $R_4^+$ versus experiment is not an artifact of the specific treatment of the statically screened Coulomb interaction in the Mo $4d$ manifold.

\section{Stability of LaMoO$_3$ \label{sec:lmo_stab}}

\begin{figure}[htp]
    \centering
    \includegraphics[width=\linewidth]{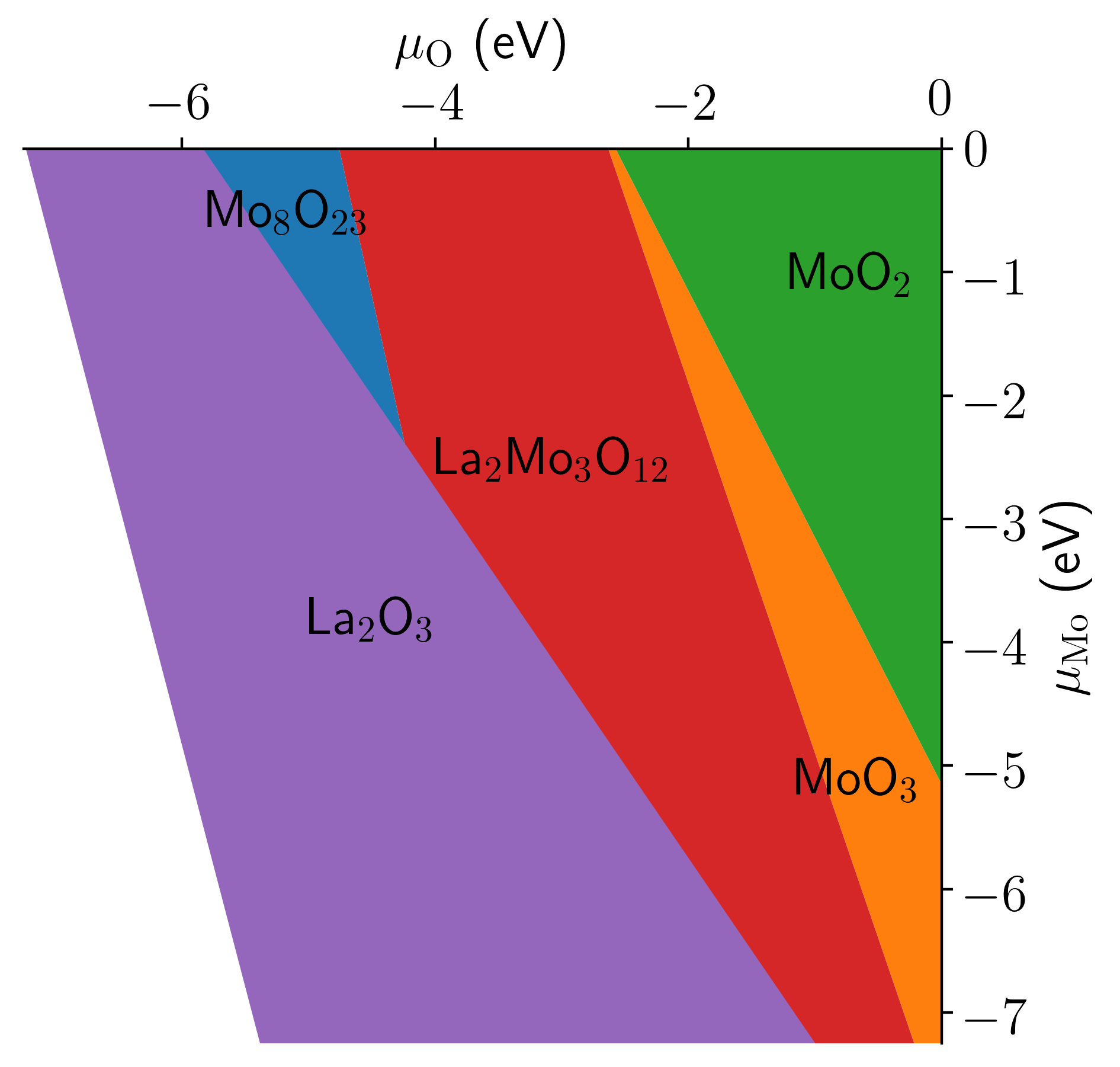}
    \caption{Stability diagram for LaMoO$_3$ with variation of the chemical potentials of oxygen ($\mu_{\text{O}}$) and molybdenum ($\mu_{\text{Mo}}$). Regions of stable phases are indicated by different colors. We do not find a stable region for LaMoO$_3$. }\label{fig:stability_diagram}
\end{figure}

Motivated by the lack of successful experimental synthesis of LMO, we determine it's thermodynamic stability towards decomposition into phases of different stoichiometry. To do this, we construct a stability diagram (Fig.~\ref{fig:stability_diagram}) for LMO with respect to the chemical potentials of oxygen $\mu_{\text{O}}$ and molybdenum $\mu_{\text{Mo}}$. The lowest-energy AFM-C $Pnma$ $U = 3$~eV structure calculated in this work was used for the energy of LMO, and all other structures were taken from Materials Project~\cite{Jain2013}. We find that LMO is not thermodynamically stable, as there is no region of allowed chemical potentials where it is the most stable stoichiometry. Ref~\onlinecite{Figen2014} reports synthesis of La$_2$Mo$_3$O$_{12}$ which we find to be stable under certain conditions (third from the right on figure). While LMO is not stable in bulk, it may be stabilized in thin-film form when grown on a perovskite substrate. 

\bibliography{biliography}

\end{document}